\documentclass{amsart}

\usepackage{booktabs}

\usepackage{graphicx}
\usepackage{subcaption}
\usepackage{wrapfig}

\usepackage{listings}
\usepackage{color}

\definecolor{dkgreen}{rgb}{0,0.6,0}
\definecolor{gray}{rgb}{0.5,0.5,0.5}
\definecolor{mauve}{rgb}{0.58,0,0.82}

\theoremstyle{definition}

\theoremstyle{remark}

\numberwithin{equation}{section}

\newcommand{\abs}[1]{\lvert#1\rvert}

\begin{document}

\title[Improved Time Warp Edit Distance]{Improved Time Warp Edit Distance \linebreak A Parallel Dynamic Program in Linear Memory}

\author{Garrett Wright}
\address{Gestalt Group LLC, Yardley, PA, 19067}
\email{garrett@gestaltgp.com}



\date{May 10, 2020}

\dedicatory{This paper is dedicated to the PhD Octopus.}


\begin{abstract}
Edit Distance is a classic family of dynamic programming problems, among which Time Warp Edit Distance refines the problem with the notion of a metric and temporal elasticity.  A novel Improved Time Warp Edit Distance algorithm that is both massively parallelizable and requiring only linear storage is presented.  This method uses the procession of a three diagonal band to cover the original dynamic program space.  Every element of the diagonal update can be computed in parallel.  The core method is a feature of the TWED Longest Common Subsequence data dependence and is applicable to dynamic programs that share similar band subproblem structure.  The algorithm has been implemented as a CUDA C library with Python bindings. Speedups for challenging problems are phenomenal.
\end{abstract}

\maketitle

\section*{Background}
Time Warping is a collection of techniques to programmatically solve the general problem of aligning time series towards superposition through a sequence of edits. Time Warp Edit Distance (TWED) is such a method distinguished by computing a proper metric in the process described by Marteau \cite{Marteau2009}.  Yielding a metric is advantageous in the contexts of machine learning and data science, particularly where work is often exploratory or experimental inquiries across large datasets.  Having a metric provides for more meaningful comparisons following from the triangle inequality.  Unfortunately, most time warping methods, including Time Warp Edit Distance, are implemented as dynamic programs which require $O(n^2)$ time and space for input time series on the order of length $n$.  In practice, this can be a limiting factor in the usefulness of such methods.

Some attempts have been made at solving Time Warp problems using approximations to reduce the computational and memory complexity.  The most popular of which appears to still require $O(n^2)$ storage on disk, but is linear in RAM during run time.  This helps to compute problems that would not fit outside of hi-memory systems, or at all, but is otherwise unsatisfying.  Recently Gold and Sharir lowered the previously quadratic bound for the related problem of deterministic Dynamic Time Warping to $O(n^2 log log log n/ log log n)$ \cite{Gold2018}.  While this is a great theoretical result, the following chart suggests a speedup of 10x for the quadratic method would greatly surpass the more sophisticated algorithm in practice, while a speedup of 100x would absolutely demolish it.  We achieve those speedups in implementation for relevant problem sizes.

\begin{figure}
\centering
\begin{subfigure}{.5\textwidth}
  \centering
  \includegraphics[width=.8\linewidth]{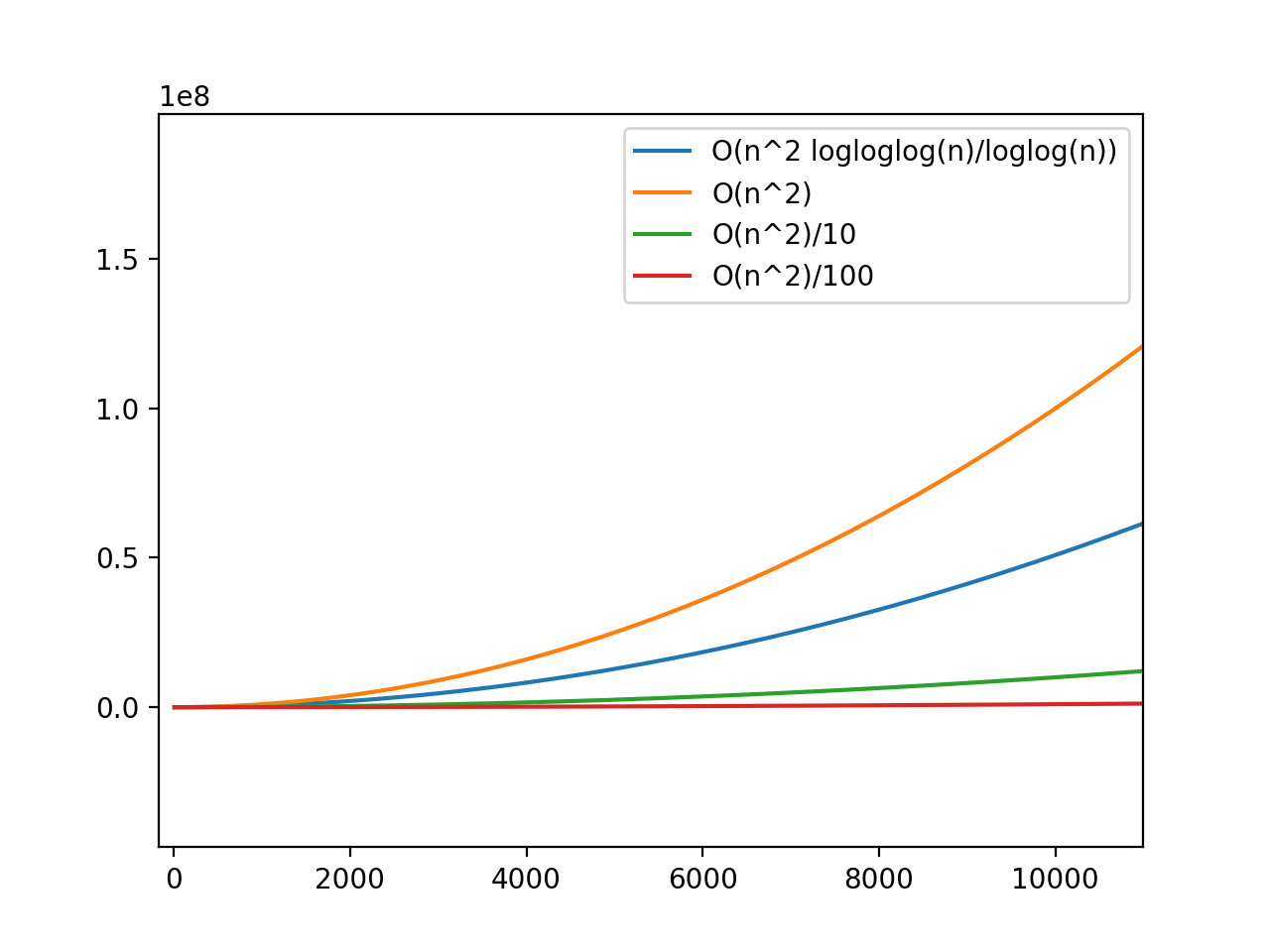}
  \caption{Up to 10K Elements}
  \label{fig:sub1}
\end{subfigure}%
\begin{subfigure}{.5\textwidth}
  \centering
  \includegraphics[width=.8\linewidth]{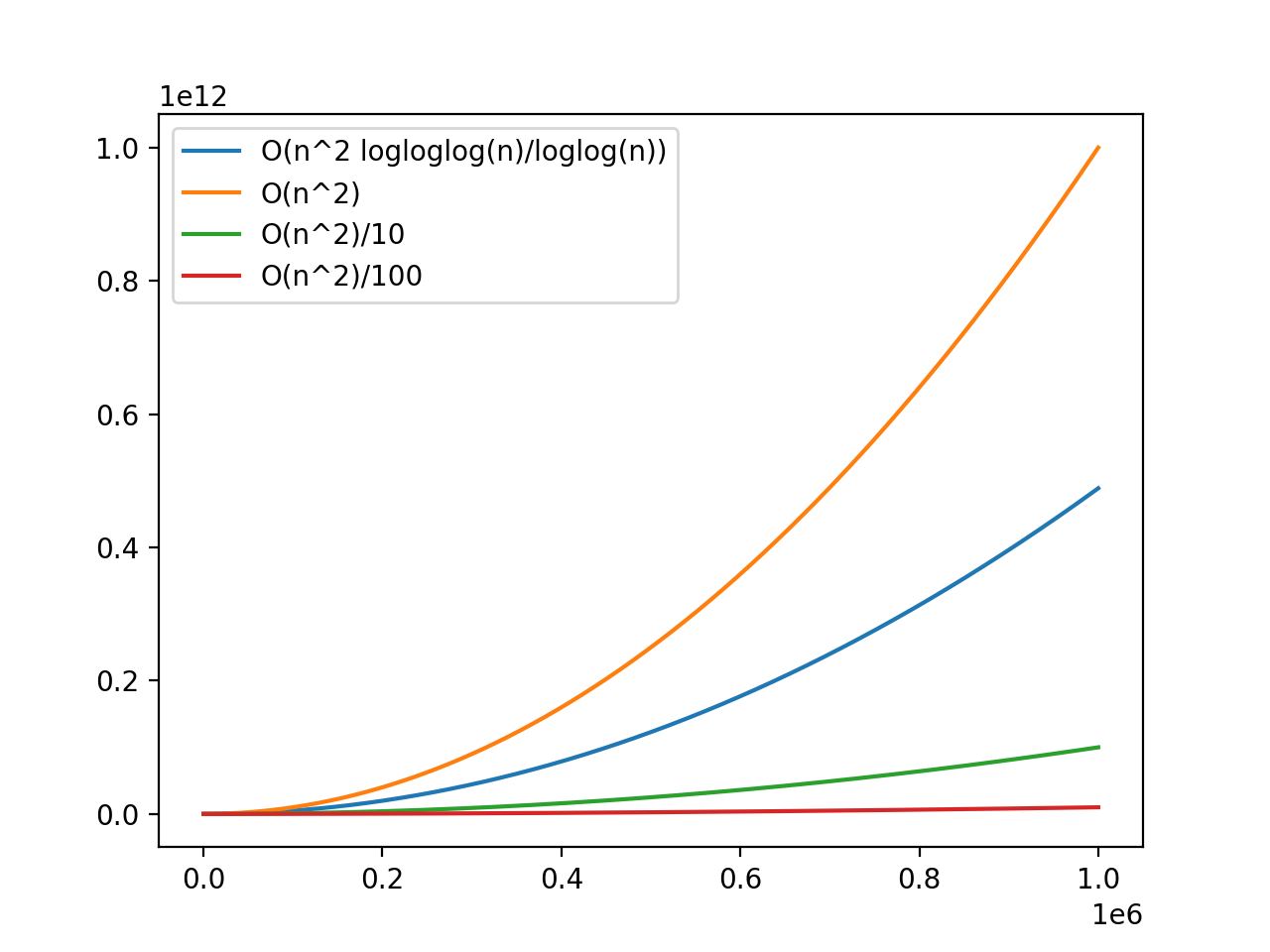}
  \caption{Up to 1M Elements}
  \label{fig:sub2}
\end{subfigure}
\caption{Complexity Scaling (ignoring constants)}
\label{fig:test}
\end{figure}

Additionally, the memory access patterns of such problems storing the complete dynamic program matrix could be considered degenerate with respect to performance during matrix updates. For instance, TWED alternates between column and row strides inside a doubly nested loop. The more sophisticated methods can have even more involved access patterns.  A key result of using the Improved Time Warp Edit Distance implementation is that the core dynamic program solver uses only three vectors of $O(n)$ length accessed in unit stride.\footnote{There are boundary conditions, but those are simply 0 or $\infty$. Their implementation and complexity-contribution can be disregarded.}

\subsection*{TWED Implementations}

I would be remiss to not mention there are several quadratic (serial) implementations of the TWED available on the web.  Particularly, TWED has its own wiki page which mentions implementations are available in C, R, Matlab, and Python.  Marteau was kind enough to freely provide his C code that is quite fast compared to the others.  This is the reference code basis.

\subsection*{A Floyd Warshall Digression}

When attempting to parallelize the TWED problem, I began by considering how other dynamic programs were parallelized (and curiously when it is possible to do correctly).  An example is the Parallel Floyd Warshall algorithm, which uses block based subproblem decomposition.  Basically, the dynamic program matrix is subdivided into blocks and it can be proven that these blocks can be correctly computed in parallel until all other related blocks are ready.\footnote{Related in the sense they are the same column or row of blocks} Then an information exchange occurs vertically and horizontally across blocks in the matrix. However, the entries in the Floyd Warshall dynamic program matrix have a graph adjacency structure, rows and columns are defined by graph vertices.  This implicitly defines the subproblem data dependence relation as totally horizontal and vertical.

\begin{figure}
   \includegraphics[width=0.5\textwidth]{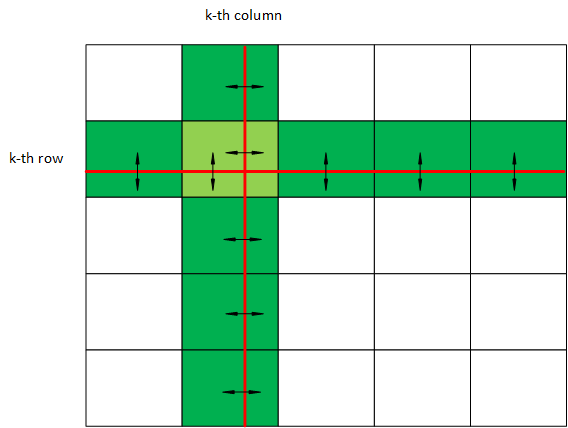}
   \caption{PFW Graphic shamelessly borrowed from Wikipedia}
\end{figure}

Assuming a large problem, PFW and similarly structured algorithms can be parallelized across many cores, but with increased communication and synchronization expense.  This also requires nontrivial programming to orchestrate the communication patterns, typically MPI.  In TWED we have a totally different data dependence, which can naturally resolve to unit size subproblems without communication or synchronization penalties.  It could be exploited with naive methods more typical of embarrassingly parallel problems, though I will use CUDA.  Exploiting this data dependence is the core result of Improved Time Warp Edit Distance.

\section*{Algorithms}

Here we review the TWED algorithm.  We will then dive into the dynamic program update step with enough detail to highlight the data dependence.  The Improved algorithm will then be detailed.

\subsection*{Time Warp Edit Distance}

We begin TWED with input arrays $A$ and $B$ which are a time series values in $R^N$.  We are also given respective real timestamps $TA$ and $TB$ corresponding to the time series values.  We will assume that we have $nA$, $nB$ samples in $A$ and $B$ respectively, where $nA$ and $nB$ are on the order of $n$ for the sake of time complexity arguments.  I will ignore the parameters $nu$ and $lambda$ here, as they do not change the mechanics of the dynamic program, and are implemented the same as in Marteau's work.\footnote{You may think of the sketch as $nu=1$ and $lambda=0$}  A $degree$ for the internal $norm$ calls is also required and typically set to $2$; we will simply refer to the operation as $norm$ and ignore the degree for this discussion.\footnote{Variable names, slightly untraditional, were chosen to match the code.}

Given the option to make edits along the path of the time series, we desire the minimal aggregate edit distance cost that can be found to align the pair of time series.  Every potential time warp edit has a computed cost that takes into account local timestamps and algorithm tuning parameters. This is described in detail by \cite{Marteau2009}. Some implementations also return the actual resultant path, but this can readily be found via backtracking or simply stored after basic observation in the forward TWED problem; it is ignored here.

\vspace{5mm} 

Sequentially:

\begin{enumerate}
	\item Compute $O(n)$ $norm$ distances in $A$ where each distance is defined as:\footnote{Note, when $i=0$ we take $A[-1]=B[-1]=0$.}
  $$DA[i] = norm(A[i] -A[i-1]) \quad \forall i \in [0,n]$$
	\item Compute $O(n)$ $norm$ distances in $B$:\footnotemark[\value{footnote}]
  $$DB[i] = norm(B[i] -B[i-1]) \quad \forall i \in [0,n]$$
	\item Initialize the dynamic program cost matrix boundary.  Assign the first row ($DP[0][:]$) and first column ($DP[:][0]$) to $\infty$.  Assign the first element, $DP[0][0]=0.$
	\item Initialize the interior of the dynamic program matrix.
	$$DP[i][j] = norm(DA[i-1]-DB[j-1]) \quad \forall i,j \in [1,n]$$
	$$DP[i][j]  \mathrel{+}= norm(DA[i-2]-DB[j-2]) \quad \forall i,j \in [2,n]$$
	\item Execute the dynamic program updates.  This is a doubly nested loop, beginning at $DP[row=1][col=1]$, 
	\begin{enumerate}
		\item We compute the cost of the following update cases:\footnote{Again, indexes outside the boundary are taken to be 0 vectors.}
		$$ delete_a = DA[row] + DP[row-1][col] + \abs{TA[row]-TA[row-1]}$$
		$$ delete_b = DB[rcol] + DP[row][col-1] + \abs{TB[col]-TB[col-1]}$$
		$$ match =  DP[row-1][col-1] + \abs{TA[row]-TB[row]} + $$ 
		$$ \abs{TA[row-1]-TB[row-1]}$$
		\item Assign $DP[row][col]=min(delete_a, delete_b, match)$
		\item This continues for the remaining columns in the row, and then again for each row in order until the end of the matrix.
	\end{enumerate}
	\item When you reach the end of the matrix, the result of the dynamic program is stored in $DP[nA][nB]$.
\end{enumerate}

\subsection*{Playing In The Band}

Some folks store in columns, and others store in rows. I don't store in either, but I know it comes out right.   In the brief description of the TWED dynamic program, the update step (5) has the doubly nested for loop. If we inspect what is happening for an element somewhere in the middle of the program, we observe the following:

\begin{itemize}
  \item To compute $delete_a$ we must look back in $A$ (one row).
  \item To compute $delete_b$ we must look back in $B$ (one col).
  \item To compute a $match$ we must look back in $A$ and $B$, one row and one column.
  \item Typical to dynamic programs, once a subproblem is optimally computed, we move on, never to update that entry again.
\end{itemize}

For the following diagrams the trivial boundary will be very lightly shaded, with dark entries representing updates and two lighter shades of blocks representing dependencies.  We'll start with the loosest bound, tighten it up, and build out an alternative storage approach.

\begin{figure}[h]
   \includegraphics[width=0.5\textwidth]{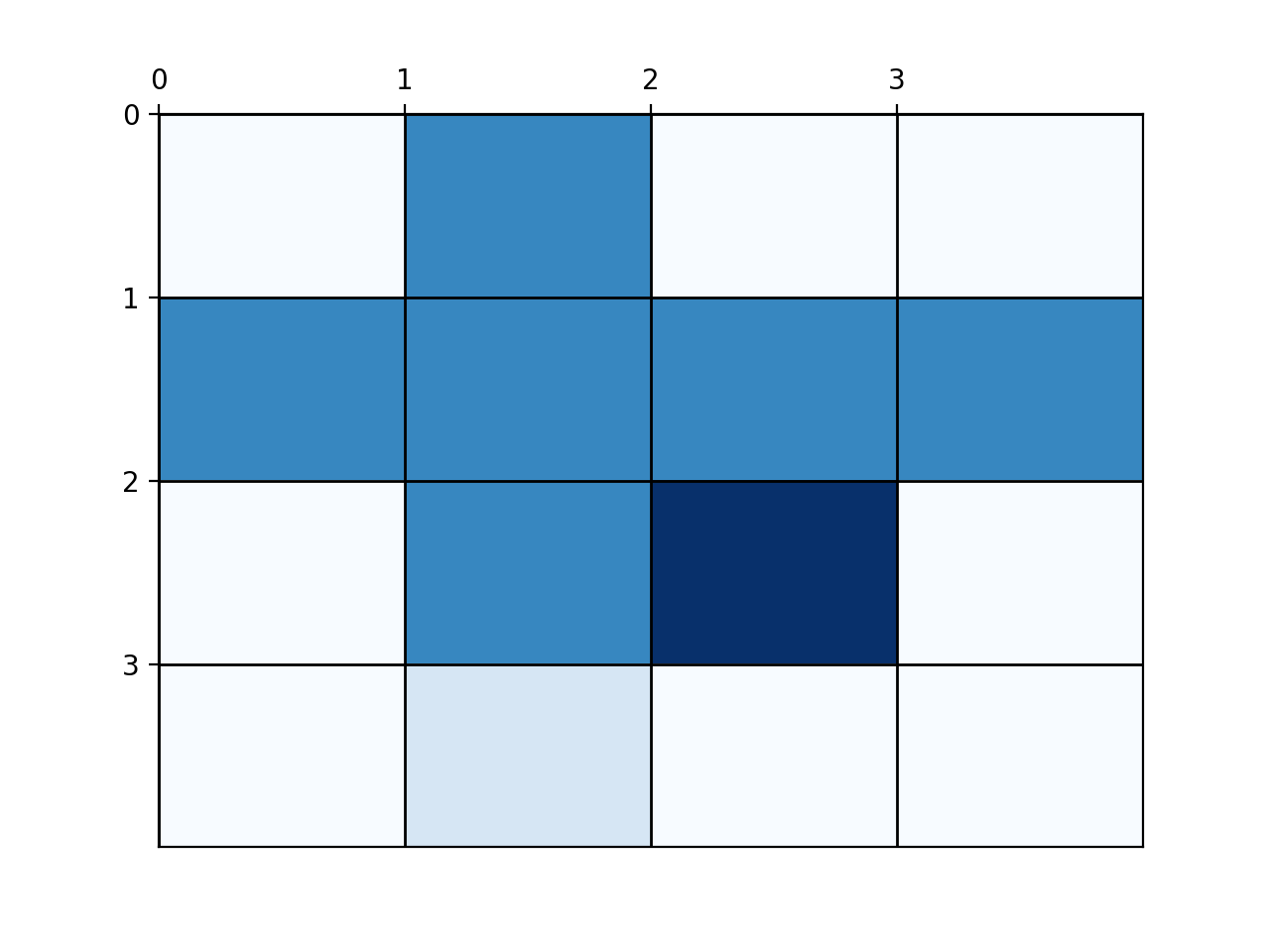}
   \caption{Crude notion of data dependence.}
\end{figure}

The boundary conditions are assigned before we begin. The problem works through the matrix in C order.  The first interior row is trivially satisfied above by the boundary. While we are computing updates moving along columns in the first interior row, data dependencies are always satisfied, because the prior column value is always the most up to date subproblem along that row. When we reach the end of the first interior row, we begin the next knowing we have the completely solved prior row, and so the entries above are the most up to date subproblems along that column. This is a key difference from Floyd Warshall, because moving along TWED rows and columns have an implied order, time.  Compare this to FW's graph vertices, which have no natural ordering; they require knowledge of entire columns and rows to complete a subproblem.

TWED can't naively compute a whole row in parallel, because the columns sequentially depend on each other.  We cannot make effective sub blocks because they would also be sequentially dependent with only the first sub block having valid initial conditions.  Similarly, we cannot compute a parallel column because the row dependence is again sequential.  Compare this to the Parallel Floyd Warshall, where we can actually compute the shortest paths that are locally true inside the domain of sub block, then exchange with other blocks globally.

With localization in mind, let's refine the observation further, and tighten up to the minimum of what we really need:

\begin{itemize}
  \item To compute $delete_a$ we must look back exactly one row \emph{in the same column}.
  \item To compute $delete_b$ we must look back exactly one col \emph{in the same row}.
  \item To compute a $match$ we must look back \emph{exactly one row and one column}.
  \item Typical to dynamic programs, once a subproblem update is computed, we move on, never to return.
\end{itemize}


So for an individual update, we actually need very few entries in the $DP$ matrix.  The matrix is initialized at $(0,0)$ $(1,0)$ and $(0,1)$, so we know there is trivially enough information to compute the first real update, entry $(1,1)$.  I'll shade the trivial boundary, and continue to use the color scheme where dark entries represent updates and lighter blocks their dependencies.

Instead of only marching on along the row, as in the original algorithm, let's consider what information \textit{we have}, and what \textit{we can compute}.  It appears we have enough information to compute two elements, the next in this row, $DP[1][2]$ as the original TWED would naturally compute and also the update $DP[2][1]$. See \ref{fig:sub1} and \ref{fig:sub2}.




\begin{figure}[h]
\centering
\begin{subfigure}{.5\textwidth}
  \centering
  \includegraphics[width=.8\linewidth]{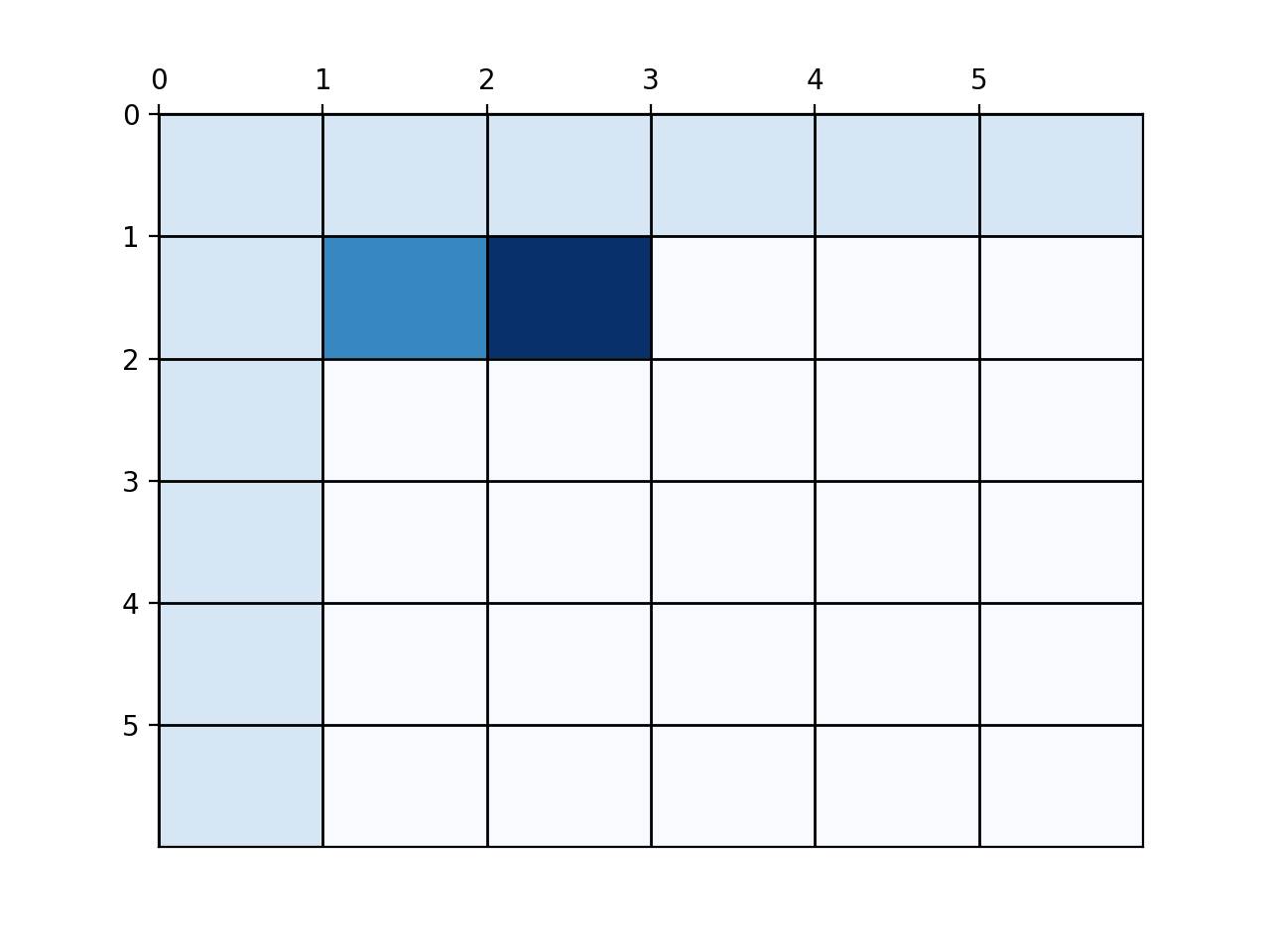}
  \caption{Update $DP[1][2]$}
  \label{fig:sub1}
\end{subfigure}%
\begin{subfigure}{.5\textwidth}
  \centering
  \includegraphics[width=.8\linewidth]{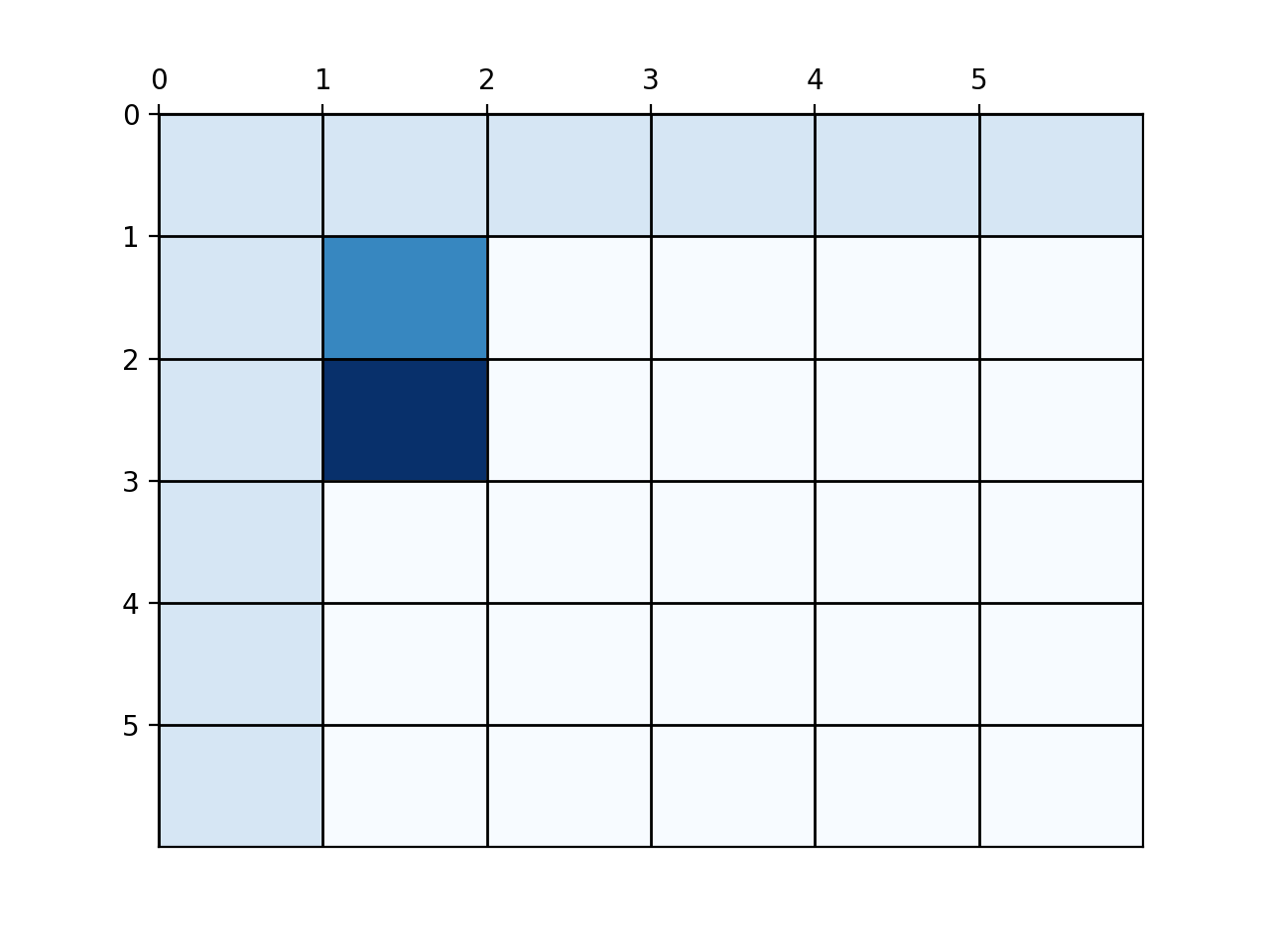}
  \caption{Update $DP[2][1]$}
  \label{fig:sub2}
\end{subfigure}
\caption{Possible to compute both two updates.}
\label{fig:test}
\end{figure}

While these share a data dependence, they are otherwise independent at this step, and they can be safely computed and assigned in parallel.

%

Let's assume we have computed both $DP[1][2]$ and $DP[2][1]$, forming a left-handed diagonal in $DP$.  Note that these diagonal entries are orthogonal to the entries mathematicians canonically refer to as diagonals.  I would like to call them ortho-diagonals, or left-handed diagonals, but for the purpose of this paper I will simply use diagonal for brevity.  Now looking forward, we consider that we can compute the items one to the right and one below any we've already computed.  Thus we can compute an entire diagonal, containing $DP[1][3]$ , $DP[2][2]$, $DP[3][1]$.\footnote{we can ignore the boundary since it is precomputed.} Again after examining their data dependencies, these entries can safely be computed and assigned in parallel. 

\begin{figure}[h]
   \includegraphics[width=0.5\textwidth]{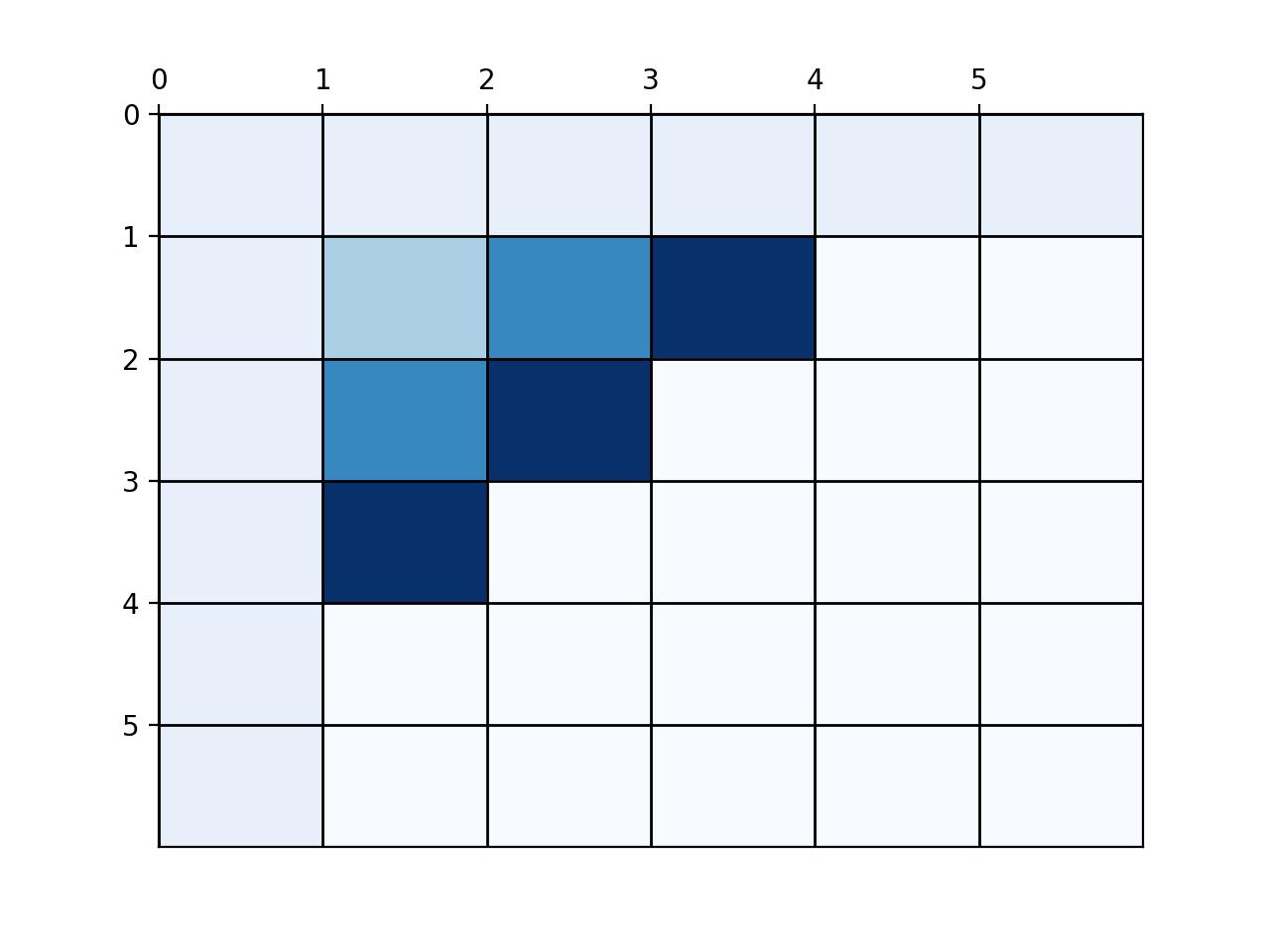}
   \caption{Diagonal Four}
\end{figure}

In fact, it is at this stage we can begin to see that given a band of the current and two most recent sequential diagonals, we can always compute the next diagonal, and that this computation can be performed in parallel.  To formalize this, let's account for the diagonals with a zero based index, starting from entry $[0,0]$.  

\begin{table}[h]
\centering
\caption{Mapping traditional indices to an Ortho-Diagonal.}
\subcaption*{Note Row+Col sum to the corresponding diagonal.}
\label{tab:mapping}
\begin{tabular}{@{}lllll@{}}
\toprule
0   & 1   & 2   & ... & z      \\ \midrule
0,0 & 0,1 & 0,2 &     & 0,z    \\
    & 1,0 & 1,1 &     & 1,z-1  \\
    &     & 2,0 &     & ...    \\
    &     &     &     & z-1, 1 \\
    &     &     &     & z,0    \\ \bottomrule
\end{tabular}
\end{table}


We can also observe, for example, that this is that last time $DP[1][1]$ will ever be referenced. This information is now \emph{packed} into the current updates. We can forget it if we want (and we should).  The fifth diagonal will go on to require the entire fourth diagonal, along with most of the third; no further prior diagonal is required now.  All that information is packed into updates already, never to be referenced directly again.

\begin{figure}[h]
   \includegraphics[width=0.5\textwidth]{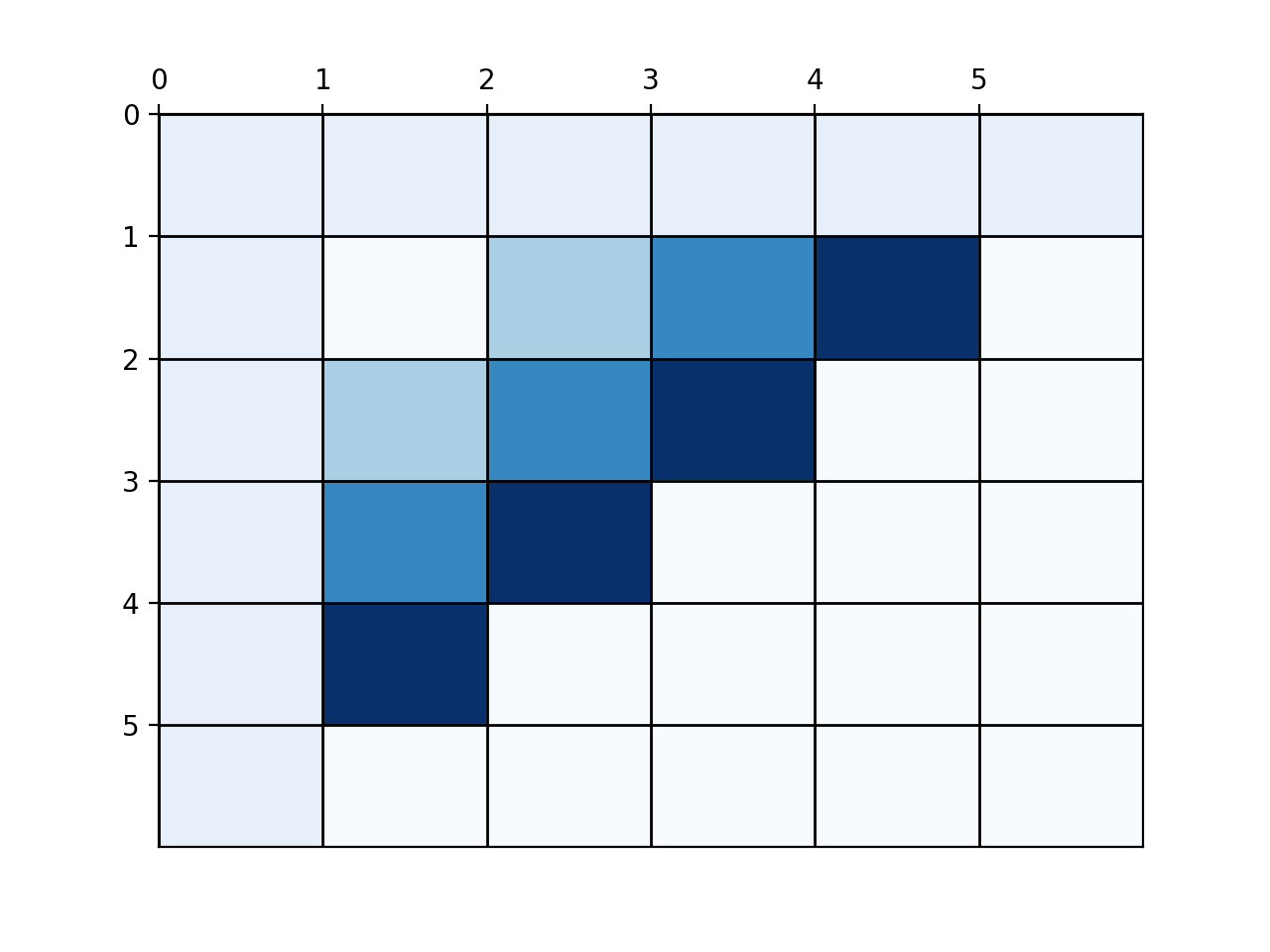}
   \caption{Diagonal Five}
\end{figure}

Given the accounting scheme in the table \ref{tab:mapping}, let's construct two utility maps between the traditional matrix entries (row, col), and my diagonal storage.\footnote{Again, boundary conditions are handled in the implementation, ignored here to keep the definitions simpler. I encourage you to play with the indices a little.  Rectangular matrices are cute...}  We want to compute which diagonal we are on, and also an index into that diagonal, which I choose to start at the lower left corner. I use the following map, though other schemes could be defined.  I was torn between this chosen scheme, and another one which required a smaller diagonal array but was buried in logic.  I chose the simpler one.
\begin{equation} \label{eq:1}
OrthoDiagonal : (row, col) \mapsto (orthodiag = row + col, idx = col)
\end{equation}
\begin{equation} \label{eq:2}
RowCol : (orthodiag, idx) \mapsto (row = orthodiag - idx, col = idx)
\end{equation}

Computing the length of the widest diagonal is the same result from classical rectangular matrices diagonals $min(row, col)$.  However, to use the basic map above we will instead let our diagonal storage hold $row + col$ elements.  Since storage space is no longer a practical concern for this algorithm, I much prefer this simpler index scheme.  The reader is welcome to consider the more challenging mapping functions to use the $min(row, col)$ sized diagonal if they wanted to implement their own code.  As a practical matter, I think the extra logic would hurt performance.

\begin{wrapfigure}[8]{l}{0.29\textwidth}
   \includegraphics[width=0.3\textwidth]{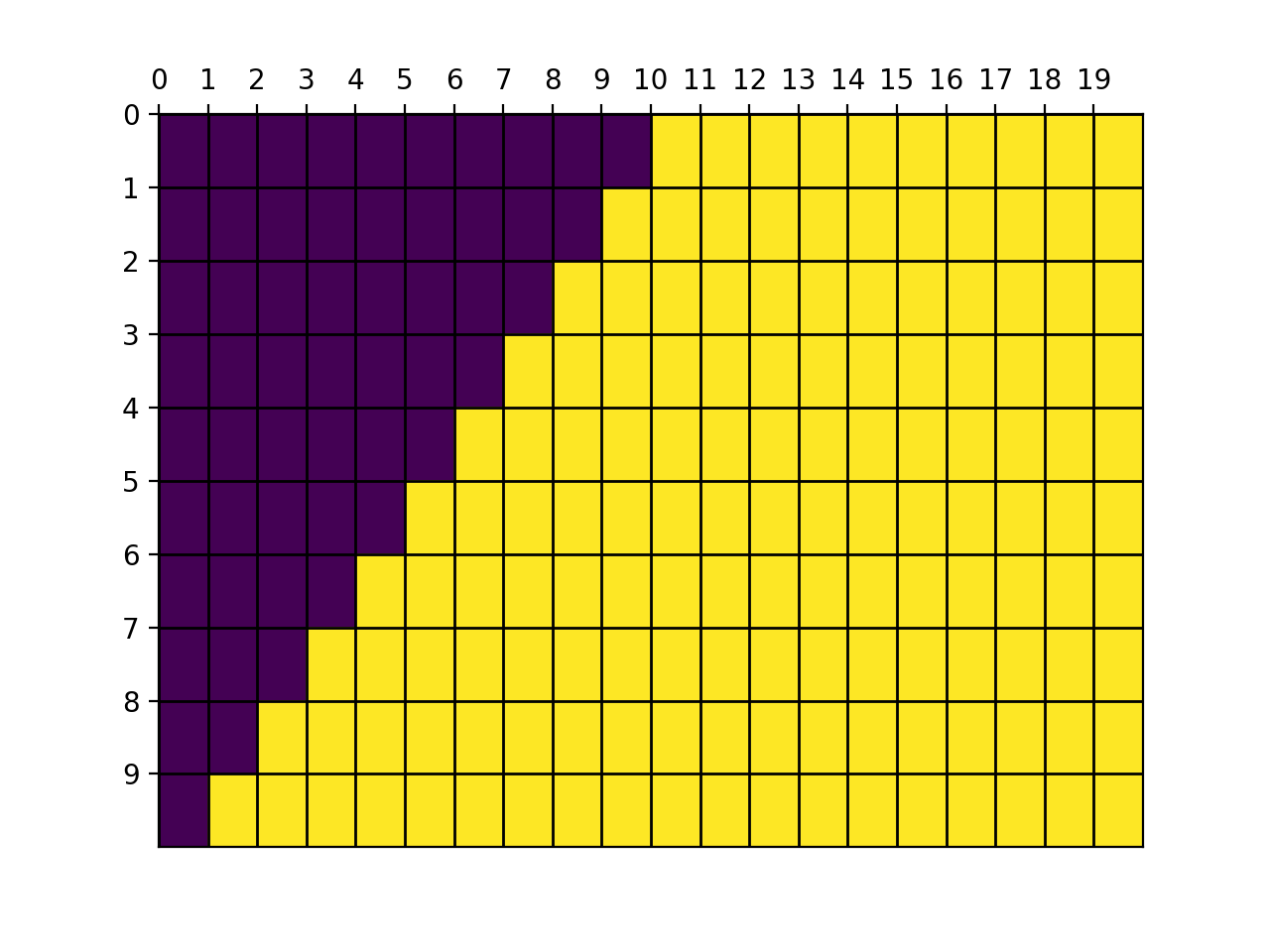}
   \caption{Covering}
\end{wrapfigure}

We can also observe that to cover a matrix uses $rows+columns-1$ diagonals. To do so by, start at [0,0], proceed down the zero column covering the upper left triangle of matrix entries with left-handed diagonals.  When you reach the lower left corner, the diagonals corresponding to nA rows are done, and we continue to proceed covering along the nB columns, but starting at column=1 (hence the columns-1) to avoid double counting the main diagonal.  This is rather simple, and I bring it up mainly because it is actually the update procession in the ITWED implementation.

Finally, we should observe that if we stack our diagonals with the correct initial offset,
all data dependencies are stored with unit stride.  This is helpful on all architectures, but 
particularly critical when it comes to a CUDA implementation.  The following graphic is a representation of the three sequences of data lookups that will occur.  The dark blue is the current active update, and the lighter blue are the data being referenced for the calculation.  This is another way to think of what will happen when computing the fifth otho-diagonal updates in the figure above.  With the exception of out of bounds checks, we achieve perfect unit stride.

\begin{figure}[h]
   \includegraphics[width=0.9\textwidth]{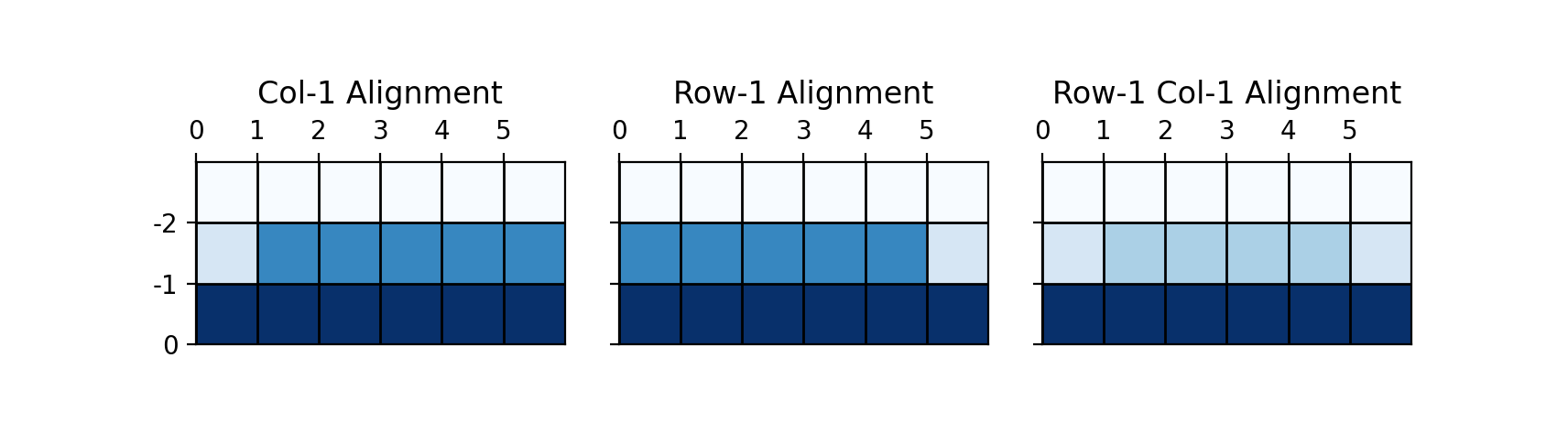}
   \caption{Coalesced alignment of data reads}
\end{figure}

\subsection*{Improved Time Warp Edit Distance}

To recap results from our adventure \textit{Playing In The Band}, we will consider a full DP matrix that has $rows = nA +1$ and $columns= nB+1$:
\begin{itemize}
	\item For the $d^{th}$ diagonal update:
	\begin{itemize}
		\item We require only the current ($z_d$), one lagged ($z_{d-1}$),  and two lagged ($z_{d-2}$) diagonals.
		\item All entries in the $z_d$ diagonal can safely be computed in parallel
	\end{itemize}
	\item The largest diagonal is $min(nA+1, nB+1)$, but we choose to store $(nA+1) + (nB+1)$ per diagonal because it simplifies indexing.
	\item For this DP matrix there are exactly $nA + nB +1$ diagonals.\footnote{$(nA +1) + (nB +1) - 1$}
\end{itemize}

\vspace{5mm} 

We can now describe the Improved Time Warp Edit Distance algorithm. Let $\parallel\forall$ to mean parallelizable for-all.

\vspace{5mm} 

\begin{enumerate}
	\item Concurrently:
	\begin{itemize}
		\item Compute $O(n)$ $norm$ distances in $A$ where each distance is defined as:\footnote{Note, when $i=0$ we take $A[-1]=B[-1]=0$.}
  $$DA[i] = norm(A[i] -A[i-1]) \quad \forall i \in [0,n]$$
		\item Compute $O(n)$ $norm$ distances in $B$:\footnotemark[\value{footnote}]
	\end{itemize}
  $$DB[i] = norm(B[i] -B[i-1]) \quad \forall i \in [0,n]$$
	\item for $d \in [1, nA+nB]$: 
	\begin{enumerate}
		\item Allocate $z_d$  with $(nA+1) + (nB+1)$ elements.
		\item $\parallel\forall$ $idx \in [0,d]$
		\begin{enumerate}	
			\item map to the respective $row$ $col$ index in a full DP matrix using \ref{eq:2}
			\item Compute initial cost on the fly:
			\begin{itemize}
				\item If $row==0$ and $col==0$, this point is the initial (minimal) point, and $d=0$
				\item Else if the $row==0$ or $col==0$, this point is a boundary, and $d=\infty$
				\item Else, we are in the interior of the dynamic program matrix:
					\linebreak\hfill $d = norm(DA[i-1] - DB[j-1]) \quad : i,j \in [1,n]$
					\linebreak\hfill $d \mathrel{+}= norm(DA[i-2] - DB[j-2]) \quad : i,j \in [2,n]$
			\end{itemize}
			\item Before we compute updates we require some index arithmetic to map between the locations in a full DP matrix and our diagonal storage.  Using \ref{eq:1} yields:
				$$idrm1 = OrthDiag(row-1, col).idx$$
				$$idcm1 = OrthDiag(row, col-1).idx$$
				$$idrm1cm1 = OrthDiag(row-1, col-1).idx$$
			\item Execute the dynamic program updates by computing the following update cases.\footnote{Yet again, indexes outside the boundary are taken to be 0 vectors.}  Note we use three diagonal arrays, the current, once lagged, and twice lagged, called $z$ $z_{-1}$ and $z_{-2}$ respectively.
				$$ delete_a = DA[row] + z_{d-1}[idrm1] + \abs{TA[row]-TA[row-1]}$$
				$$ delete_b = DB[col] + z_{d-1}[idcm1] + \abs{TB[col]-TB[col-1]}$$
				$$ match =  z_{d-2}[idrm1cm1] + \abs{TA[row]-TB[row]} + $$ 
				$$ \abs{TA[row-1]-TB[row-1]}$$
			\item  Assign result $z_{d}[idx]=min(delete_a, delete_b, match)$.
		\end{enumerate}
		\item If $z>2$: Free $z_{d-2}$
	\end{enumerate}
	\item When you complete all diagonals, the result of the dynamic program is stored in the final $z_d$ at $OrthDiag(nA,nB).idx$.
\end{enumerate}

At any given point we have at most three diagonals in memory, each using $(nA+1) + (nB+1)$ memory.  We additionally must store both input time series, their $norm$ distances, and the timestamps. All these are trivially $O(n)$ space.

Computationally we are still quadratic number of steps. To cover the full DP matrix, for each $nA+nB+1$ diagonal we compute at most $min(nA+1, nB+1)$ updates.  On the other hand, because we can now parallelize over all the diagonal entries, and our memory is stored with unit stride, we can effectively leverage thousands of cores which would yield $O(n^2/p)$ for $p$ cores.  The CUDA programming model naturally exposes several thousand GPU cores, and this brings us to discuss cuTWED.

\section*{cuTWED Implementation Remarks}

Largely the CUDA code cuTWED is taken from the ITWED algorithm directly, with the addition of managing corner cases and boundary conditions.  Because we have taken care to understand the data dependence and have stored our diagonal data in a way that is accessed with unit stride, we can simply map the parallel for in ITWED (2b) to a 1D grid of 1D blocks following the standard CUDA programming model.

To avoid extraneous malloc (3a) and free (3c) calls, cuTWED manage's its own memory by manipulating pointers in a cycle:

\begin{lstlisting}
    tmp_ptr = DP_diag_lag_2;
    DP_diag_lag_2 = DP_diag_lag;
    DP_diag_lag = DP_diag;
    DP_diag = tmp_ptr;
\end{lstlisting}

Using CUDA streams cuTWED is able to squeeze out some additional concurrency from the device, by computing algorithm steps in (1) concurrently.

Further, two distance kernels are provided, and automatically selected at run time based on the dimension of the time series inputs.  Recall the $lp$-norm formula for ${\displaystyle \mathbf {x} =(x_{1},\ldots ,x_{n})}$:
$$\left\|\mathbf {x} \right\|_{p}:={\bigg (}\sum _{i=1}^{n}\left|x_{i}\right|^{p}{\bigg )}^{1/p}$$

Given a time series of high dimensional vectors, it can be advantageous to use extra parallelism available on the device to compute all the $pow$ calls, storing into fast shared memory simultaneously. These can then be accumulated quickly from shared memory, then the nth root taken is to complete the norm.  Power calls are notoriously expensive, essentially being several instructions, and while aggregate register pressure should be considered, generally it is effective to perform the $pow$ operations in parallel on the GPU \cite{cuda2013best}. For example, given an input time series in $R^{32}$, we can compute raising all vector elements simultaneously with little overhead.  I am not sure the speedup merits the code complexity, and this optimization may be removed in the future.

The current release of cuTWED implements a batch mode for large system of time series.  Given two lists of time series (potentially the same) as large arrays, we can use  two dimensions in CUDA to process multiple time series and multiple diagonal entries as a 2D grid of 2D blocks. The output of batch mode is a distance matrix corresponding to all pairs of entries in the two lists of time series. A slightly optimized unreleased version makes more use of streaming concurrency overlaps.

In a future release of cuTWED it is planned to implement some light auto-tuning logic, and capability to harness multiple GPUs.  Time depending it may be optimized further

\section*{Performance}

In cases where the problem was computable by twed.c\footnote{Implementations using the other higher level languages, while convenient, are orders slower still and are not considered here.}, cuTWED is demonstrably two orders of magnitude faster on current GPU hardware\footnote{Intel Xeon E5-2680v4 with Nvidia P100}.  Even on older hardware\footnote{Intel i7-2600K CPU with TitanZ Kepler GPU}, more typical of personal computers than a research facility, great speedups, above a single order of magnitude are attained.

\subsection*{Basic Methodology and Results}
The reference twed.c code is compiled ($-O2$) into a shared library and bound with a minimal lightweight Python binding using the python `cffi` library. The comparison methodology is to generate a repeatable sequence of random time series of particular sizes (powers of two in this case).  Timing results from the batches are averaged together, though this probably wasn't necessary in hindsight because the timings were pretty stable if the machines were otherwise unloaded.

\begin{figure}[b]
\centering
\begin{subfigure}{.5\textwidth}
  \centering
  \includegraphics[width=.8\linewidth]{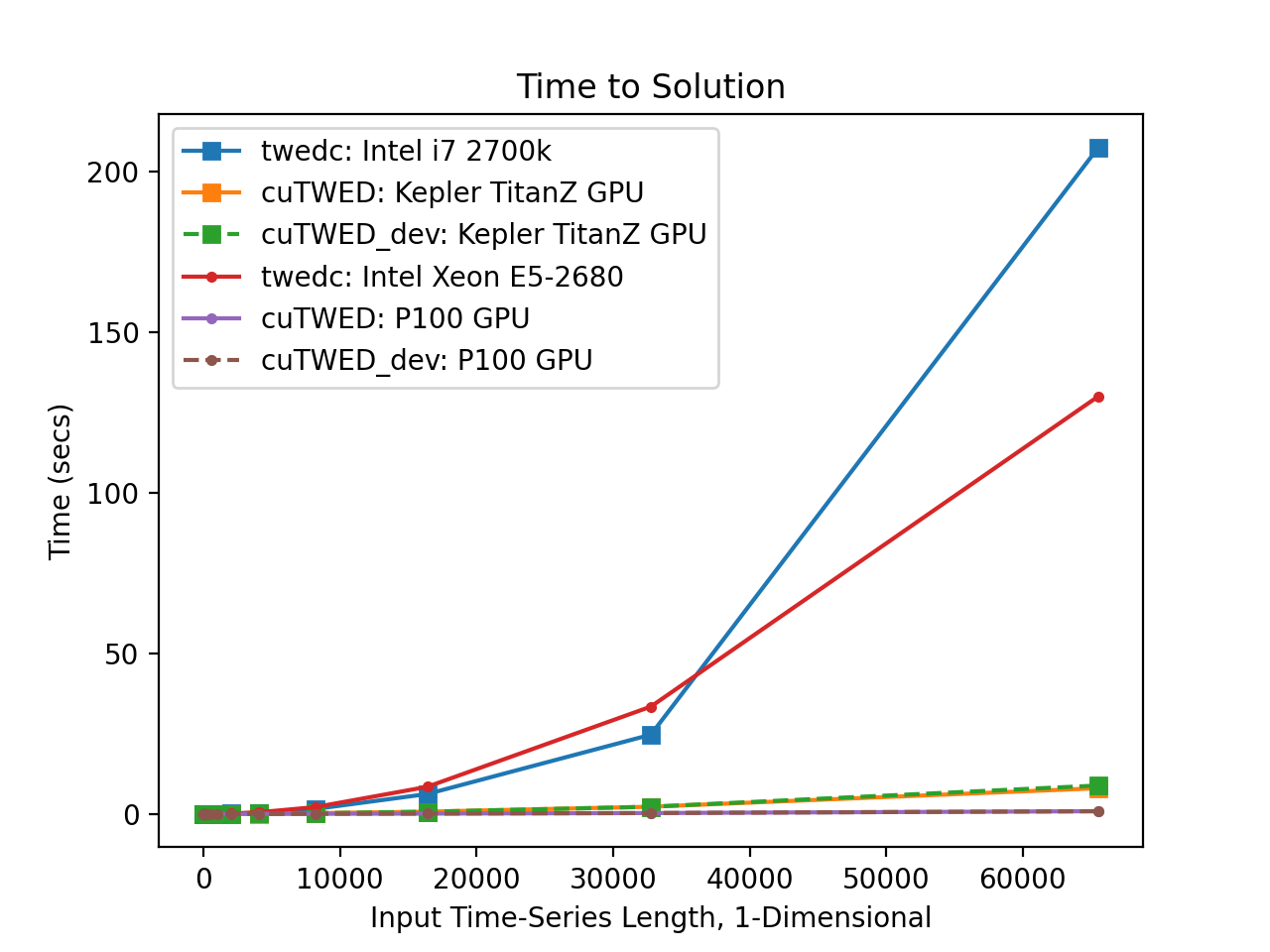}
  \caption{Walltime}
  \label{timing:sub1}
\end{subfigure}%
\begin{subfigure}{.5\textwidth}
  \centering
  \includegraphics[width=.8\linewidth]{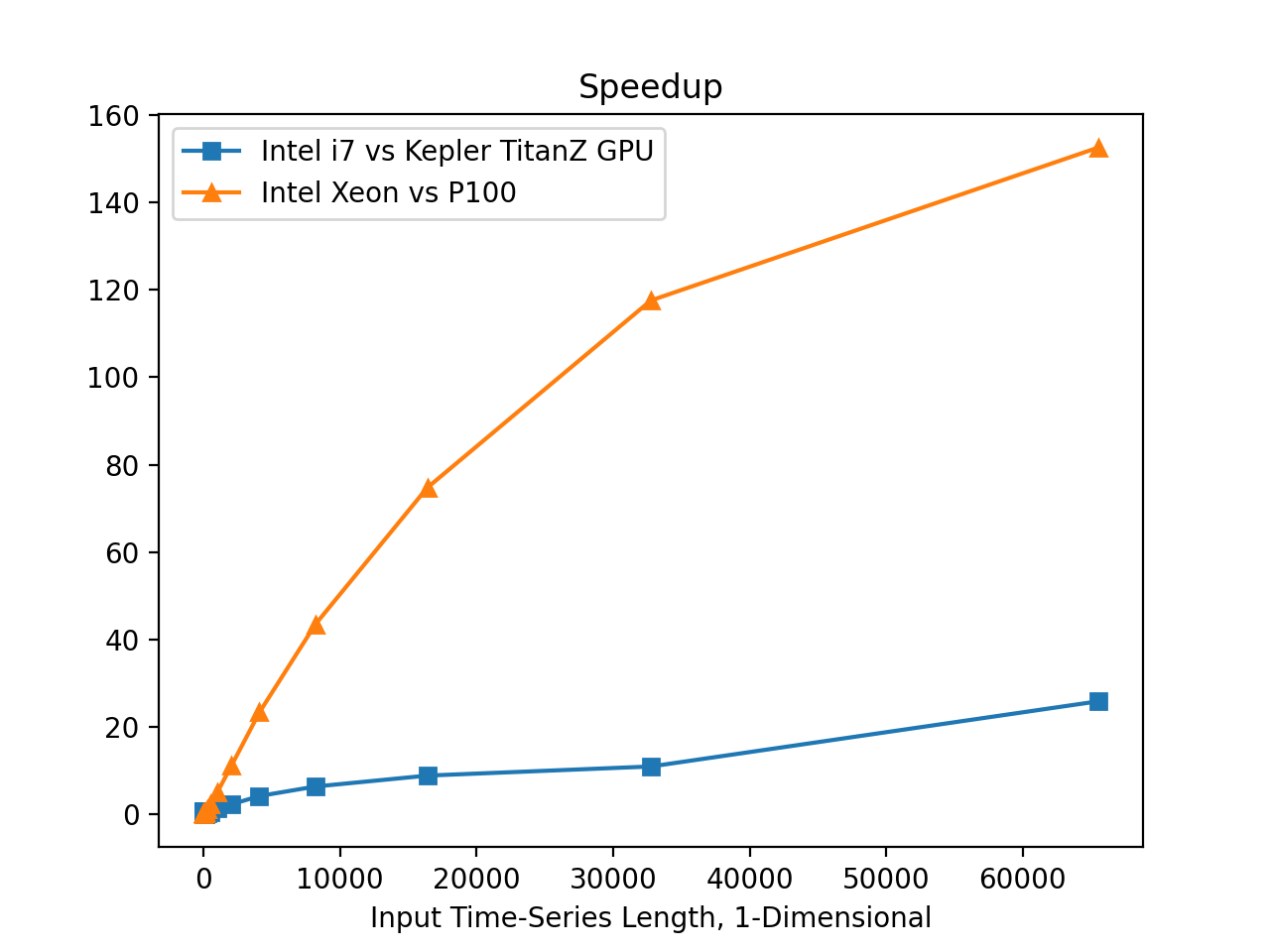}
  \caption{Relative Speedup}
  \label{timing:sub2}
\end{subfigure}
\caption{Timings}
\label{fig:test}
\end{figure}

For cuTWED we feed it the same sequence of time series, first measuring times for a managed $cuTWED.twed$ call from a host.  We repeat the batched experiments for $cuTWED.twed\_dev$ which is intended for problems where the data resides on GPU already.  From this we can also extrapolate a practical measure of host to device transfer cost overhead, which is trivial\footnote{Overhead is generally less than machine noise for this problem.}.  This should make sense because the input data is linear, we don't preprocess it, and our algorithm is quadratic.  Walltime is plotted in figure \ref{timing:sub1}.

All of the experiments were performed in double precision.  While single precision will obviously reduce storage by half, it had marginal performance benefit otherwise. Any gain would probably be lost to required casting overhead in practice, should you not already be in singles. The author recommends using the precision your data provided in.

Times are listed in a traditional table at the end of the paper for reference, see Table \ref{tab:raw-times-table}.



 Perhaps a more telling perspective is relative speedup on the two test systems plotted in figure  \ref{timing:sub2}.  The two test systems consist of one late model desktop with a GPU circa 2014, comparable to desktops and laptops many people might have personally. The second system is a common configuration for production or research facilities with current hardware.  \emph{We achieve triple digit speedups on current hardware.}  Over 150x speedup and still climbing when the host code is no longer able to run.


Because we can, I've run this out to time series of one million elements. No problem. With series this large we finally have measurable, but still trivial, transfer costs.  Extremal sized cases can be found in figure \ref{fig:sub1}. Because no original implementation can fit such problems, even with high memory nodes, there is no comparison data.



\begin{figure}[t]
\centering
\begin{subfigure}{.5\textwidth}
  \centering
  \includegraphics[width=.8\linewidth]{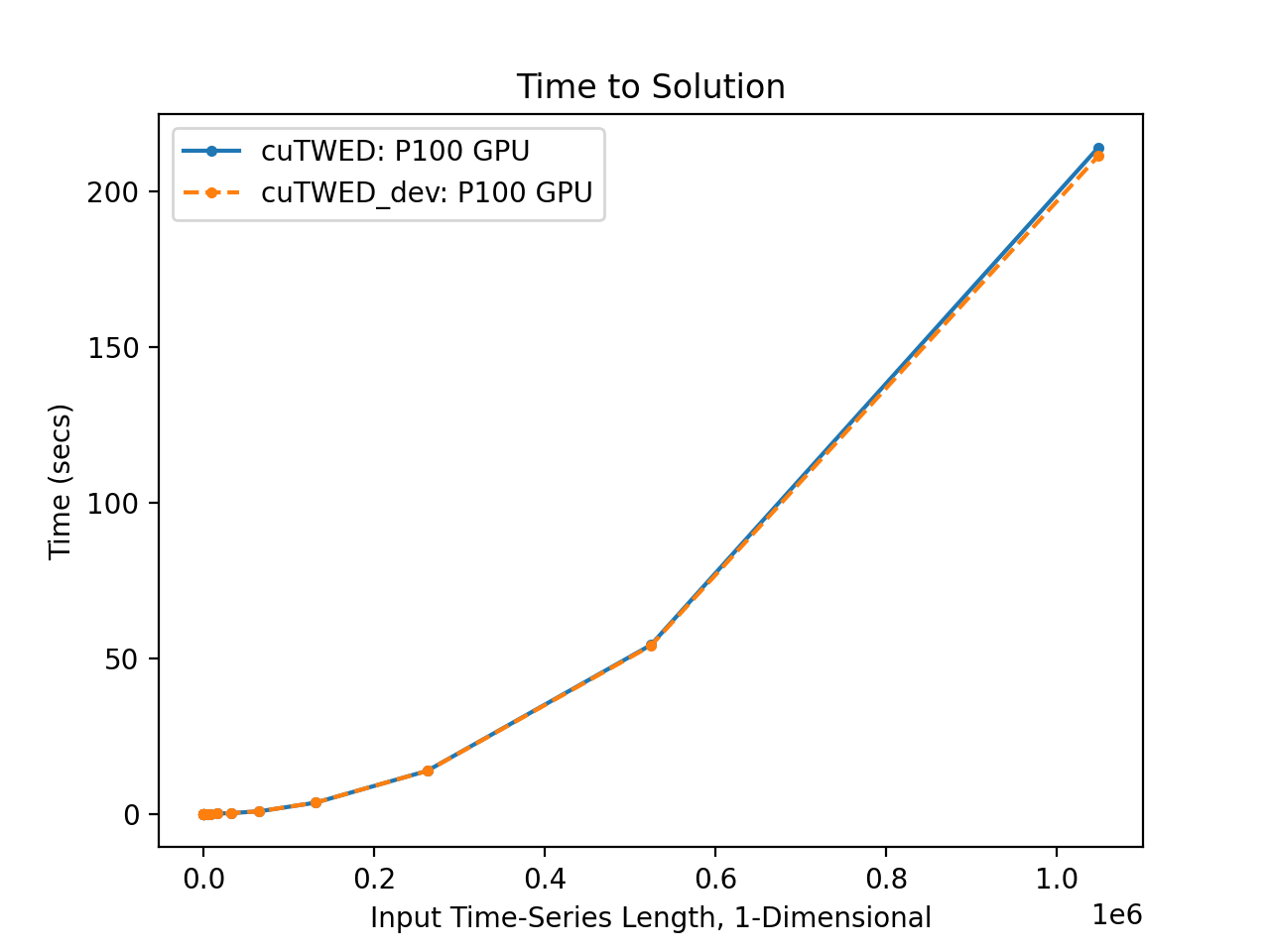}
  \caption{Walltime}
  \label{fig:sub1}
\end{subfigure}%
\begin{subfigure}{.5\textwidth}
  \centering
  \includegraphics[width=.8\linewidth]{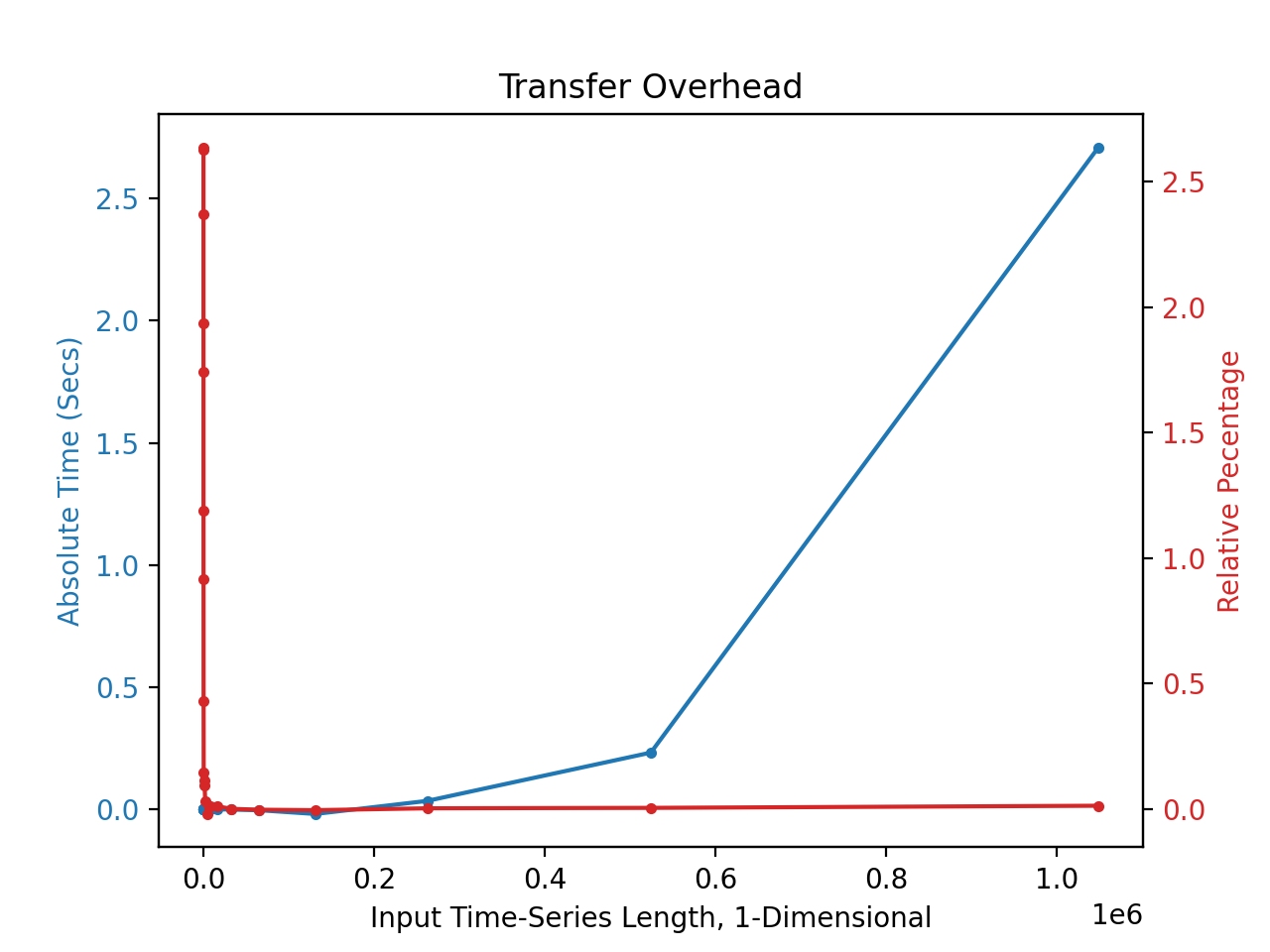}
  \caption{Overhead}
  \label{fig:sub2}
\end{subfigure}
\caption{Large Scale Timings}
\label{fig:test}
\end{figure}

\subsection*{Higher Dimensions}

The only original TWED implementation with reasonable speed (twed.c) lacks support for $R^N$ inputs. To compare such problems requires modifying the reference code to admit $R^N$.  This is straightforward, and was done using essentially the same code as in $cuTWED$. Similar experiments were repeated with increasing dimension $N$.  The use of $R^N$ didn't really have any practical time difference in either implementation. Basically, since the $R^N$ component is only used once for $A$ and $B$ in a linear computation, until $N$ is on the order of $n$ this is overshadowed by the dynamic program matrix computation. And we get to skip a chart on that.

\subsection*{Validation using Real Datasets}

\begin{figure}[h]
   \includegraphics[width=0.9\textwidth]{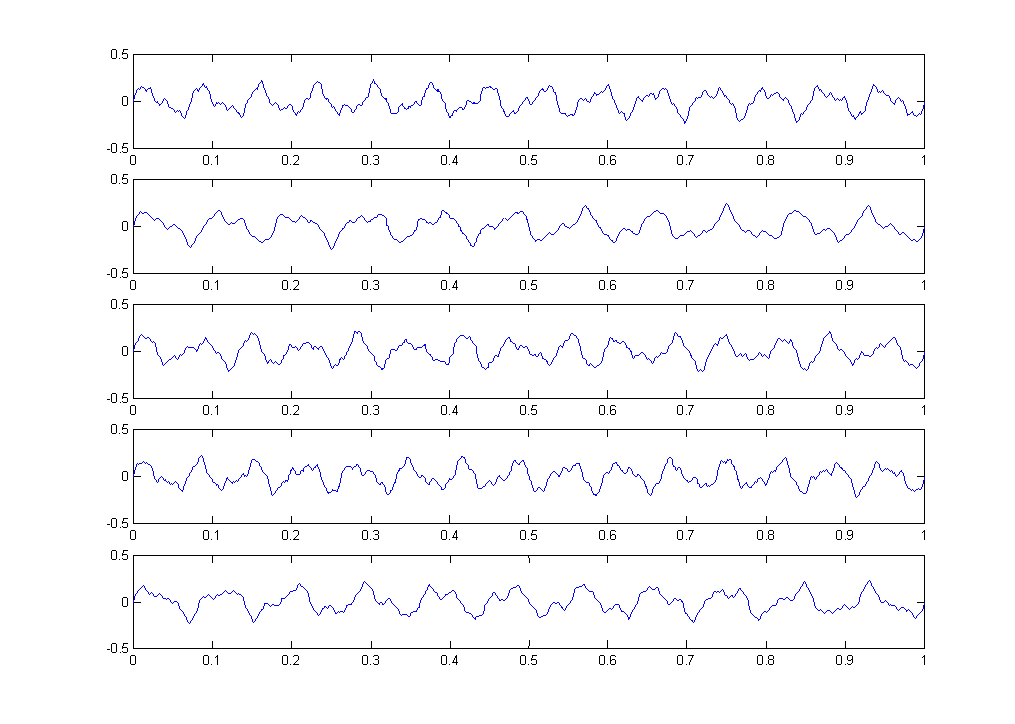}
   \caption{UCI Synthetic Control Time Series}
   \label{UCIcontrol1}
\end{figure}

\begin{figure}[h]
   \includegraphics[width=0.9\textwidth]{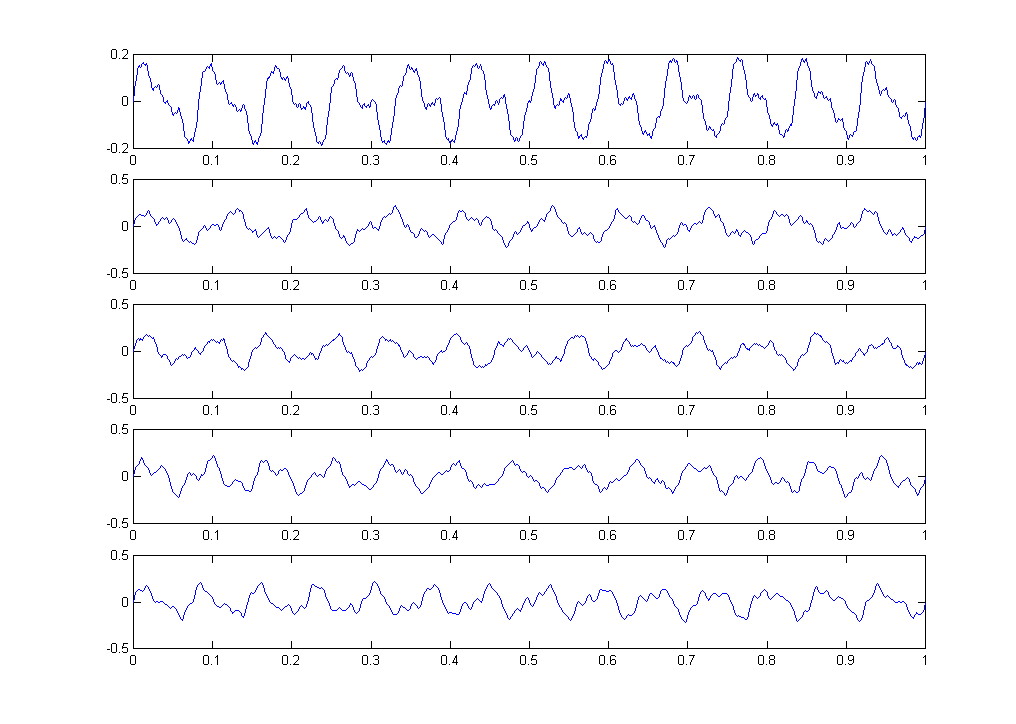}
   \caption{UCI Synthetic Control Time Series 6-10}
   \label{UCIcontrol2}
\end{figure}

\begin{figure}[h]
   \includegraphics[width=0.9\textwidth]{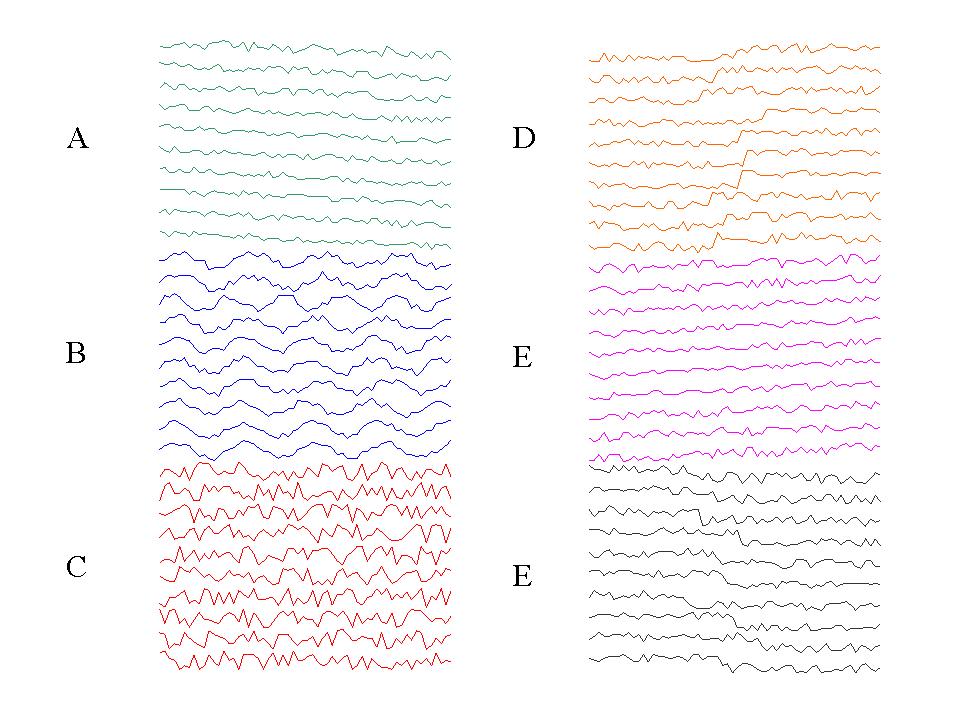}
   \caption{UCI Pseudo Periodic Time Series}
   \label{UCIPseudo}
\end{figure}

To validate and performance test using real world data, three datasets were considered.  From the UCI Machine Learning Repository \footnote{http://archive.ics.uci.edu/ml} \cite{Dua2019} two time series datasets were chosen.  The Synthetic Control Chart Time Series Data Set consists of 600 1D time series including Cyclic, Trending, and Shifting series.  The Pseudo Periodic Synthetic Time Series Data Set consists of 10 time series each with 100k points.  The pseudo periodic data does not repeat and is targeted for testing database indexing schemes.  Most importantly the results are checked to match between cuTWED and TWED.  In most cases they match to the bit, but occasionally they are about 1E-14 different in RMSE, which is consistent with the aggregated use of optimized pow/sqrt calls on different chips.  We compute all pairs of TWED using both tools and document timings.  For quick reference figures  \ref{UCIcontrol1}  \ref{UCIcontrol2} \ref{UCIPseudo} of the testing time series are found at the end of the paper.

For another larger problem we'll consider the MNIST\footnote{http://yann.lecun.com/exdb/mnist/} database of handwritten digits.  We chose to use MNIST to validate computations in $R^N$ by unraveling one dimension of the two-dimensional image as a 28 sample time axis, and treating the other as a vector in $R^{28}$.  As a timing example a random digit from the test set is selected and distances computed against the training set of 60k images.  Again we check for parity, and this query is also documented in the timings.

\begin{table}[h]
\centering
\caption{Timings From Validation Exercises (seconds)}
\label{tab:ValidationTimings}
\resizebox{\textwidth}{!}{%
\begin{tabular}{@{}lllll@{}}
\toprule
                  & TWED  & cuTWED &  Size & Speedup \\ \midrule
Synthetic Control & 22    & 1    & 600x600 x 60 $R^1$ & 22x     \\
Pseudo Periodic   & 15750\footnotemark & 147  & 10x10 x 100k $R^1$ & 107x    \\
MNIST  Single Query      &    30   &  0.20   & 1x60k x 28  $R^{28}$ &   150x  \\
MNIST All Query      &  -    &  $<$2.6hr  & 60kx60k x 28  $R^{28}$ & - \\  \bottomrule
\end{tabular}%
}
\end{table}

\footnotetext{Required use of my 192GB high memory server, one allocation of the dynamic program matrix is 80GB in doubles. The same experiment can easily be computed with cuTWED on a common late model GPU capable desktop or laptop in a matter of minutes.}

Note the distance matrix is symmetric, and currently the cuTWED batch call computes the complete distance matrix.  The nested loop I use for the CPU reference code is only computing the upper triangle (roughly half).  When this optimization is added to the CUDA code I would expect the above batch time speedups to roughly double.

\section*{Generalized Dynamic Programming Extensions}

\subsection*{Direct Extension: LCS}

Problems with similar banded optimal subproblem may benefit from this approach. For example the length of the classic Longest Common Subsequence can be solved with my diagonal band trick in linear memory.  Simply note the fact that your data dependence is totally local, just as in cuTWED.  For an LCS subproblem you need priors up and to the left, and the initial sequences.  We've covered how to map from any diagonal band back to rows and columns, ie the original sequences.  Given this, you can compute the update value. This can be performed, in parallel no less, and repeated.  When you reach the end of the dynamic programming matrix, you have computed the LCS length.  If you would like the actual subsequence(s) this will require two lagged diagonals, same as we've discussed.  The band of diagonals will keep enough information to determine if there is a new optimum and what the right-handed diagonal parent entry is.  By bookkeeping the path of parent entries we are forward constructing the same path as in the backtracking LCS method. There is the case of multiple LCS, which is simply additional bookkeeping of a path for each sequence.

I believe there is also a way to use the band technique with the Matrix Chain Multiplication ordering optimization. However, that particular problem is commonly optimized with a hash table in place of the complete dynamic program matrix, and I suspect because of this the practical application would not be improved with my technique.

\subsection*{Indirect Extension}

Other problems may not immediately present as banded, but could be reinterpreted with some consideration.  One might desire to do this to either reduce memory footprint or better exploit the GPU.  Because the banded storage is stride ideal in linear memory, such diagonal banded algorithms may prove more effective in time to solution over others with the same or better theoretical computational time.  Also if the banded storage is correct for an application, some parallelized problems might benefit from a total reduction in communication overhead naturally provided by this technique.

\section*{Summary}
A novel technique requiring only linear memory to solve a family of dynamic programs has been described.  Additionally, cuTWED, an open-source high-performance CUDA and Python library using the technique has been published under GPL.  Timings presenting one to two orders of magnitude speedups, while already outstanding, are known to have at least another factor of two left in optimizations.  The technique and software was similarly validated against the original TWED algorithm using three classic machine learning datasets. 

\qed

\vspace{5mm} 

	The author would like to thank Dr Igor Rivin for his encouragement to write this down for others, instead of tossing it in the piles of my other unfinished work once I figured it out. I hope others can make some use of it.

\vfill

\bibliographystyle{amsplain}
\bibliography{references}

\providecommand{\bysame}{\leavevmode\hbox to3em{\hrulefill}\thinspace}
\providecommand{\MR}{\relax\ifhmode\unskip\space\fi MR }
\providecommand{\MRhref}[2]{%
  \href{http://www.ams.org/mathscinet-getitem?mr=#1}{#2}
}
\providecommand{\href}[2]{#2}
\begin{thebibliography}{1}

\bibitem{cuda2013best}
C~Cuda, \emph{Best practice guide, 2013}, 2013.

\bibitem{Dua2019}
Dheeru Dua and Casey Graff, \emph{{UCI} machine learning repository}, 2017.

\bibitem{Gold2018}
Omer Gold and Micha Sharir, \emph{Dynamic time warping and geometric edit
  distance: Breaking the quadratic barrier}, ACM Trans. Algorithms \textbf{14}
  (2018), no.~4.

\bibitem{Marteau2009}
P.~{Marteau}, \emph{Time warp edit distance with stiffness adjustment for time
  series matching}, IEEE Transactions on Pattern Analysis and Machine
  Intelligence \textbf{31} (2009), no.~2, 306--318.

\end{thebibliography}

\begin{table}[htbp]
\centering
\caption{Raw Time to Solution}
\label{tab:raw-times-table}
\resizebox{\textwidth}{!}{%
\begin{tabular}{@{}lllll@{}}
\toprule
N          	& cuTWED	& cuTWED\_dev	& twedc	& Speedup  \\ \midrule
1048576 	& 214.22 	& 211.51	& & \\
524288  	& 54.317 	& 54.084	&		\\
262144  	& 13.863	& 13.827 	&		\\
131072  	& 3.6166	& 3.6344 	&		\\
65536   	& 0.8545  	& 0.8571	& 129.91	& 152x \\
32768   	& 0.2852 	& 0.2851 	& 33.400	&  117x \\
16384   	& 0.1152 	& 0.1140  	& 8.4386	&  73x \\
8192    	& 0.0504	& 0.0499  	& 2.1325	&  42x \\
4096    	& 0.0238	& 0.0243  	& 0.5439	&  23x \\
2048    	& 0.0122	& 0.0118 	& 0.1300	&  11x \\
1024    	& 0.0067	& 0.0059  	& 0.0301	& 4.5x \\
\bottomrule
\end{tabular}%
}
\end{table}

\end{document}